\documentclass[prl,twocolumn,amsmath,amssymb,floatfix]{revtex4-2}
\usepackage{graphicx}
\usepackage{dcolumn}
\newcolumntype{d}[1]{D{.}{.}{#1}}
\usepackage{bm}
\usepackage{longtable}
\usepackage{mathtools}

\usepackage{epstopdf}

\begin{document}

\title{Hexagonal-to-base-centered-orthorhombic $\bm{4Q}$
  charge density wave order in kagome metals KV$_3$Sb$_5$,
  RbV$_3$Sb$_5$, and CsV$_3$Sb$_5$}

\author{Alaska Subedi} 

\affiliation{CPHT, CNRS, Ecole Polytechnique, IP Paris, F-91128
  Palaiseau, France} 

\date{\today}

\begin{abstract}

I search for the ground state structures of the kagome metals
KV$_3$Sb$_5$, RbV$_3$Sb$_5$, and CsV$_3$Sb$_5$ using first principles
calculations.  Group-theoretical analysis shows that there are
seventeen different distortions that are possible due to the phonon
instabilities at the $M$ $(\frac{1}{2},0,0)$ and $L$
$(\frac{1}{2},0,\frac{1}{2})$ points in the Brilouin zone of the
parent $P6/mmm$ phase of these materials.  I generated these
structures for the three compounds and performed full structural
relaxations that minimize the atomic forces and lattice stresses.  I
find that the $Fmmm$ phase with the order parameter $M_1^+$ $(a,0,0)$
$+$ $L_2^-$ $(0,b,b)$ has the lowest energy among these possibilities
in all three compounds.  However, the $Fmmm$ exhibits a dynamical
instability at its $Z$ $(0,0,1)$ point, which corresponds to the $A$
$(0,0,\frac{1}{2})$ point in the parent $P6/mmm$ phase.  Condensation of
this instability leads to a base-centered orthorhombic structure with
the space group $Cmcm$ and $4Q$ order parameter $M_1^+$ $(a,0,0)$
$+$ $L_2^-$ $(0,b,b)$ $+$ $A_6^+$
$(\frac{1}{2}c,\frac{-\sqrt{3}}{2}c)$.

\end{abstract}


\maketitle

\section{Introduction}

KV$_3$Sb$_5$, RbV$_3$Sb$_5$, and CsV$_3$Sb$_5$ are members of a new
family of kagome metals recently discovered by Ortiz \textit{et al.}
\cite{orti19}.  These materials occur in a layered structure with the
hexagonal space group $P6/mmm$.  The kagome layers in these materials
are made up of corner-shared V triangles interlaced with a hexagonal
lattice of Sb ions.  These are sandwiched by honeycomb layers of Sb
ions such that the V ions are situated inside a face-shared lattice of
Sb octahedra, which are then alternately stacked with hexagonal layers of
$A$ ( = K, Rb, or Cs) ions to form the full three-dimensional
structure.

First principles calculations predict that these materials host
electronic bands that are both linearly dispersive and flat depending
on the direction in the reciprocal space, and the Fermi level lies
near Dirac points that are in the vicinity of van Hove
singularities \cite{orti19}.  This electronic structure has been confirmed
by angle resolved photoelectron spectroscopy (ARPES) experiments
\cite{yang20,orti20,li21a,liu21a,wang21a,naka21,hu21a,luo21a,kang21,cho21,hu21b,lou21,luo21b}.
Calculations also show that these materials are $\mathbb{Z}_2$
topological metals with nontrivial surface crossings
\cite{orti20,orti21a}, and there is evidence of topological surface
states in ARPES measurements \cite{hu21a}.  Although neutron
scattering and muon spin resonance experiments find no evidence of
localized electronic magnetic moments or long-range magnetic order
\cite{orti19,kenn21,miel21,yu21c}, large anomalous Hall effect has
been observed in these materials \cite{yang20,yu21a}.  This has been
attributed to a chiral flux phase \cite{feng21a,yu21c}, but no such
chiral flux current was found in spin-polarized scanning tunneling
measurments \cite{li21c}.

All three compounds exhibit charge density wave (CDW) and
superconducting transitions \cite{orti20,orti21a,yin21a}.  The CDW
transition occurs at 78, 103, and 94 K, respectively, for
KV$_3$Sb$_5$, RbV$_3$Sb$_5$, and CsV$_3$Sb$_5$.  The respective
superconducting transition temperatures $T_c$ are 0.9, 0.9, and 2.5 K.
Penetration depth and nuclear magnetic resonance (NMR) experiments
find evidence for a nodeless $s$-wave superconducting gap
\cite{duan21,mu21}, but there are also experimental results that are
suggestive of unconventional pairing \cite{wang20,zhao21a,wang21d}.
Regardless, the superconducting state seems to have multiple Fermi
sheets with different gaps \cite{ni21,orti21b,fu21,xu21,gupt21}. Strain,
pressure, doping, and sample thickness experiments indicate that there
is competition between CDW order and superconductivity
\cite{zhao21a,chen21a,du21,zhan21,chen21c,tsir21,song21c,yin21b,song21a,yu21b,wang21c,qian21}.

The crystal structure of the CDW state has not been fully determined.
An early scanning tunneling microscopy (STM) experiment found an
inplane $2\times2$ superlattice reconstruction in KV$_3$Sb$_5$ that is
possibly chiral in nature \cite{jian21}. Subsequent STM experiments
have found evidences for $2\times2\times2$, as well as $1\times4$
charge orders in the three materials
\cite{zhao21b,lian21,chen21b,li21b,xu21,shum21,wang21b,li21c}. X-ray
scattering and NMR experiments support the presence of a
$2\times2\times2$ charge order \cite{li21a,song21b}, while one x-ray
diffraction study finds evidence for a $2\times2\times4$
superstructure \cite{orti21b}. The $1\times4$ superlattice modulation
that breaks the hexagonal symmetry has been argued to be a surface
phenomenon \cite{shum21,wang21b}, but the $c$-axis resistivity only
exhibits twofold symmetry in the presence of a rotating inplane magnetic
field, suggesting that the $C_6$ rotational symmetry may also be
broken within the bulk \cite{xian21}. Optical and pump-probe
spectroscopy experiments support both nesting-driven and
unconventional scenarios for the CDW formation
\cite{zhou21,uyku21a,wang21,uyku21b}.

First principles calculations by Tan \textit{et al.}\ predict that the
$2\times2$ order is caused by tri-hexagonal distortions within the V
kagome layers \cite{tan21,miao21}, whereas those by Ratcliff
\textit{et al.}\ show that $C_6$ symmetry is broken by simultaneous
condensation of the unstable phonons at the $M$ $(\frac{1}{2},0,0)$
and $L$ $(\frac{1}{2},0,\frac{1}{2})$ wave vectors \cite{ratc21}.  The
charge, orbital and superconducting instabilities in these materials
have been the subject of additional theoretical studies
\cite{feng21a,denn21,zhao21c,lin21a,wu21,park21,sett21,feng21b,lin21b,chris21,taza21,lin21c,gu21,labo21}.
Nevertheless, a detailed study that identifies the full crystal
structure of these materials in their ground state, which is necessary
to understand the mechanism underlying their CDW order and
superconductivity, is still lacking.

In this paper, I use density functional theory based first principles
calculations to search for the lowest energy structures of these
materials among the possible structures that arise due to the phonon
instabilities present in these materials.  Group-theoretical analysis
shows that there are seventeen different structural distortions
possible due to the phonon instabilities at $M$ and $L$ points in the
parent $P6/mmm$ phase of these materials.  I generated these distorted
structures for all three materials and performed full structural
relaxtions by minimizing the atomic forces and lattice stresses.  I find
that the $Fmmm$ phase with the order parameter $M_1^+$ $(a,0,0)$ $+$
$L_2^-$ $(0,b,b)$ has the lowest energy among these structures.
However, the $Fmmm$ phase exhibits a phonon instability at the $Z$
$(0,0,1)$ point in its Brillouin zone, which is equivalent to the $A$
$(0,0,\frac{1}{2})$ point of the $P6/mmm$ phase.  This instability leads to
a base-centered orthorhombic structure with the space group $Cmcm$, and
it has the $4Q$ order parameter $M_1^+$ $(a,0,0)$ $+$ $L_2^-$
$(0,b,b)$ $+$ $A_6^+$ $(\frac{1}{2}c,\frac{-\sqrt{3}}{2}c)$ with
respect to the $P6/mmm$ phase.

\section{Computational Approach}

All structural relaxation and phonon calculations presented here were
performed using the optB88-vdW exchange-correlation functional that
accurately treats the van der Waals interaction \cite{optb88}.  The
{\sc quantum espresso} package \cite{qe}, which is a pseudopotential
based planewave code, was used for the structural relaxations and the
phonon calculations of the three compounds in the $P6/mmm$ phase. I
used the pseudopotentials generated by Dal Corso \cite{dalc14} and
energy cutoffs of 60 and 600 Ry for the basis-set and charge density
expansions, respectively, in these calculations.  An $8\times8\times4$
$k$-point grid and a Marzari-Vanderbilt smearing of 0.01 Ry was used
for the Brillouin zone integration.  The dynamical matrices were
calculated on an $8\times8\times4$ $q$-point grid using density
functional perturbation theory \cite{dfpt}, and Fourier interpolation
was used to obtain the phonon dispersions.

I used the {\sc isotropy} package to enumerate all possible structural
distortions due to the phonon instabilities at $M$ and $L$
\cite{isotr}. The calculated phonon eigenvectors were then used to
generate the distorted structures on $2\times2\times2$ supercells of
all three compounds.  To minimize the consumption of computational
resources, the {\sc vasp} package was used to perform full structural
relaxations of these supercells.  A planewave cutoff of 400 eV,
$k$-point grid of $8\times8\times4$, and Methfessel-Paxton smearing
of 0.1 eV were used in these calculations.

Phonon calculations of the $Fmmm$ phase using density functional
perturbation theory was not feasible with the available computational
resources because of the large number of atoms in its primitive unit
cell.  Hence, the frozen-phonon approach as implemented in the {\sc
  phonopy} package in combination with the {\sc quantum espresso} code
as the force calculator was used to calculated the phonon dispersions
of the $Fmmm$ phase of the three materials \cite{phonopy}. A 288-atom
$2\times2\times2$ supercell and a $3\times3\times3$ $k$-point grid was
utilized in these calculations.  The frozen-phonon approach was also
used to check for phonon instabilities at $Z$ $(0,0,\frac{1}{2})$ and
$S$ $(\frac{1}{2},\frac{1}{2},0)$ points in the $Cmcm$ phase of
CsV$_3$Sb$_5$ using 144-atom $1\times1\times2$ and $2\times1\times1$
supercells, respectively.  These calculations used $k$-point grids
of $6\times6\times2$ and $3\times6\times4$, respectively.

The spin-orbit interaction was neglected in all structural relaxation
and phonon calculations.  I made extensive use of the {\sc findsym}
\cite{findsym}, {\sc amplimodes} \cite{ampli}, and {\sc spglib}
\cite{spglib} packages in the symmetry analysis of the relaxed
structures.

The electronic structure of the $P6/mmm$, $Fmmm$, $Cmcm$ phases of
KV$_3$Sb$_5$ was calculated using the generalized full-potential
method as implemented in the {\sc wien2k} package. I used the
fully-relaxed structures obtained using the optB8-vdW functional but
performed the calculations with the generalized gradient approximation
of Perdew, Burke and Ernzerhof \cite{pbe}.
Muffin-tin radii of 2.5, 2.44, and 2.45 a.u.\ were used for K, V, and
Sb, respectively.  The planewave cutoff was set by $RK_{\textrm{max}}
= 7$, where $K_{\textrm{max}}$ is the planewave cutoff and $R$ is the
smallest muffin-tin radius used in the calculations.
$32\times32\times16$, $16\times16\times16$, and $16\times16\times8$
$k$-point grids were used for the $P6/mmm$, $Fmmm$, and $Cmcm$ phases,
respectively.  The density of states was calculated using
$64\times64\times32$, $28\times28\times28$, and $24\times24\times12$
$k$-point grids, respectively.  The spin-orbit interaction was
included in these calculations.

\section{Results and Discussion}

\subsection{What is the lowest energy structure due to the
  phonon instabilities at $\bm{M}$ and $\bm{L}$?}


 
\begin{table}[htb]
  \caption{\label{tab:latt} Calculated structural parameters of
    $A$V$_3$Sb$_5$ ($A$ = K, Rb, Cs) obtained using the opt88B-vdW
    functional in the high-temperature $P6/mmm$ phase. The atomic
    coordinates are $A$ $1a$ $(0,0,0)$, V $3g$
    $(\frac{1}{2},\frac{1}{2},\frac{1}{2})$, Sb1 $1b$ $(0,0,\frac{1}{2})$,
    and Sb2 $4h$ $(\frac{2}{3},\frac{1}{3},z_{Sb})$.}
  \begin{ruledtabular}
    \begin{tabular}{l  d{1.4} d{1.4} d{1.4}  @{\hspace{1em}} d{1.4} d{1.4} d{1.4}}
      &   \multicolumn{3}{c}{theory}  & \multicolumn{3}{c}{experiment\footnote{Ref.~\cite{orti19}}} \\
      $A$   & \multicolumn{1}{c}{$a$} & \multicolumn{1}{c}{$c$}
      & \multicolumn{1}{c}{$z_{\textrm{Sb}}$} & \multicolumn{1}{c}{$a$} & \multicolumn{1}{c}{$c$}
      & \multicolumn{1}{c}{$z_{\textrm{Sb}}$} \\
      \hline
      K    &  5.4760 & 8.9370 & 0.7572  & 5.48213 & 8.95802 & 0.753 \\
      Rb   &  5.4907 & 9.0910 & 0.7517  & 5.4715  & 9.073   & 0.74984 \\
      Cs   &  5.5097 & 9.3065 & 0.7444  & 5.4949  & 9.3085  & 0.74217
    \end{tabular}
  \end{ruledtabular}
\end{table}

\begin{figure*}
  \includegraphics[width=\textwidth]{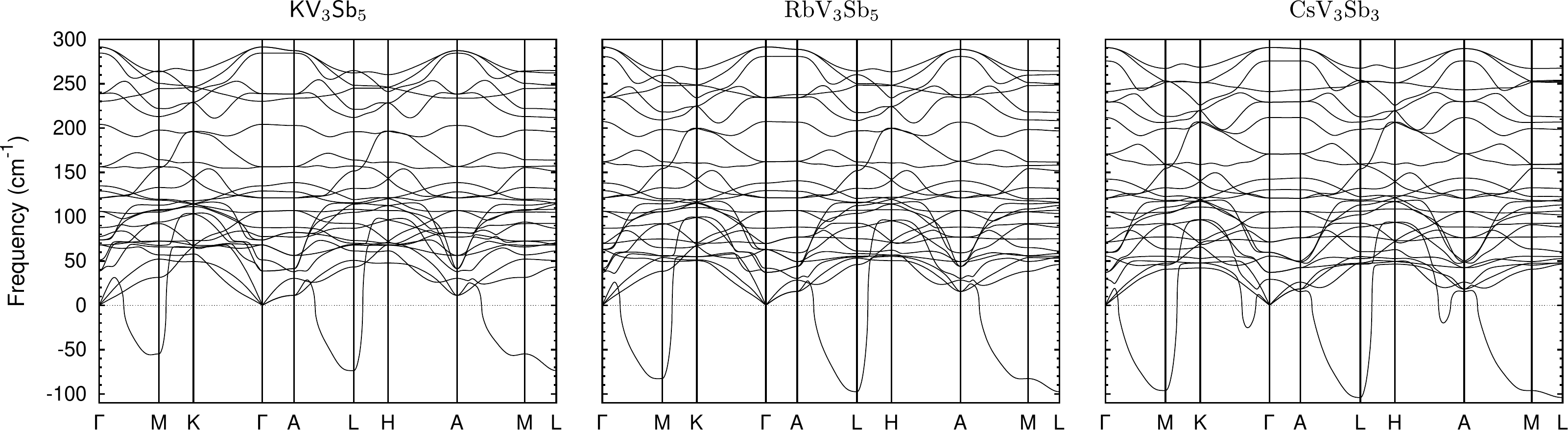}
  \caption{Calculated phonon dispersions of fully-relaxed KV$_3$Sb$_5$, RbV$_3$Sb$_5$, and 
  CsV$_3$Sb$_5$ obtained using the optB88-vdW functional show that all three materials have
  an unstable phonon branch.  The high-symmetry points are $\Gamma$ $\left(0,0,0\right)$, 
  $M$ $\left(0,\frac{1}{2},0\right)$,  $K$ $\left(\frac{1}{3}, \frac{1}{3}, 0\right)$, 
  $A$ $\left(0,0,\frac{1}{2}\right)$, $L$ $\left(0,\frac{1}{2}, \frac{1}{2}\right)$, and 
  $H$ $\left(\frac{1}{3}, \frac{1}{3}, \frac{1}{2}\right)$. The coordinates are given in terms
  of the reciprocal lattice vectors and the imaginary frequencies are indicated by negatives values.}
  \label{fig:pd}
\end{figure*}

The calculated structural parameters of KV$_3$Sb$_5$, RbV$_3$Sb$_5$,
and CsV$_3$Sb$_5$ obtained using the optB88-vdW functional in the
high-temperature $P6/mmm$ phase are given in Table~\ref{tab:latt}.
These are in good agreement with the experimentally determined values
\cite{orti19}.  The phonon dispersions of the three compounds obtained
using these calculated structures are shown in Fig.~\ref{fig:pd}, and
they agree reasonably well with previous calculations
\cite{tan21,chen21c,cho21,ratc21}.  The dispersions show a
non-degenerate branch that is unstable along the path from $M$
$(0,\frac{1}{2},0)$ to $L$ $(0,\frac{1}{2}, \frac{1}{2})$.
Additionaly, the phonon branch shows a much weaker instability near
$\Gamma$ and $A$ $(0,0,\frac{1}{2})$ in CsV$_3$Sb$_5$.  The unstable
branch continuously connects to an acoustic mode near the Brillouin
zone center.  But all acoustic branches emerge with real velocities
from the Brillouin zone center, indicating that an acoustic
instability does not cause the structural transition in these
materials.  In all three materials, the instability at $L$ is slightly
larger than at $M$.  The strength of these instabilities increase as
the atomic mass increases from K, Rb, to Cs.  This is consistent with
RbV$_3$Sb$_5$ having a higher structural transition temperature than
KV$_3$Sb$_5$, but at odds with CsV$_3$Sb$_5$ having a lower structural
transition temperature than RbV$_3$Sb$_5$.  In CsV$_3$Sb$_5$, the
phonon instability is spread to a greater extent along the
high-symmetry paths and the unstable branch is relatively flat along
$M$--$L$, which should both allow for more structrual fluctuations,
and this might explain the discrepancy.

The phonon branches in all three materials are relatively flat along
the out-of-plane directions $\Gamma$--$A$ and $M$--$L$, 
indicating that the interlayer bonding in these materials is weak.  This agrees with the lack of charge 
sharing between the alkali metal and V-Sb layers ascertained from electronic structure calculations \cite{orti19}.
The electronic structure calculations furthermore indicate that the alkali metal ions primarily act as a electron
donors, and the very similar phonon dispersions of the three compounds confirms this view.

\begin{table}
  \caption{\label{tab:disp-vec} Calculated displacement vectors of the unstable mode at $L$ 
    $(0,\frac{1}{2},\frac{1}{2})$ for $A$V$_3$Sb$_5$ ($A$ = K, Rb, Cs). The displacement vector has the dimension of 
    length and is normalized to unity. The atoms are at the positions $A$ (0,0,0), 
    V(1) $(\frac{1}{2},\frac{1}{2},\frac{1}{2})$, V(2) $(0,\frac{1}{2},\frac{1}{2})$,
    V(3) $(\frac{1}{2},0,\frac{1}{2})$, Sb(1) $(\frac{2}{3}, \frac{1}{3}, z_{\textrm{Sb}})$, 
    Sb(2) $(\frac{1}{3}, \frac{2}{3}, z_{\textrm{Sb}})$, Sb(3) $(\frac{1}{3}, \frac{2}{3}, -z_{\textrm{Sb}})$,
    Sb(4) $(\frac{2}{3}, \frac{1}{3}, -z_{\textrm{Sb}})$, and Sb(5) $(0,0,\frac{1}{2})$. }
  \begin{ruledtabular}
    \begin{tabular}{l d{2.2} d{2.2} d{2.2}  d{3.2} d{2.2} d{2.2} d{3.2} d{2.2} d{2.2}}
      &   \multicolumn{3}{c}{KV$_3$Sb$_3$} & \multicolumn{3}{c}{RbV$_3$Sb$_3$} & 
      \multicolumn{3}{c}{CsV$_3$Sb$_3$} \\
      atom   & \multicolumn{1}{c}{$x$} & \multicolumn{1}{c}{$y$}
      & \multicolumn{1}{c}{$z$} & \multicolumn{1}{c}{$x$} & \multicolumn{1}{c}{$y$}
      & \multicolumn{1}{c}{$z$} & \multicolumn{1}{c}{$x$} & \multicolumn{1}{c}{$y$}
      & \multicolumn{1}{c}{$z$} \\
      \hline
      $A$   &  0.00 & 0.00 &  0.05 &  0.00 &  0.00 &  0.02 &  0.00 &  0.00 &  0.03 \\
      V(1)  & -0.66 & 0.08 &  0.00 & -0.67 &  0.05 &  0.00 & -0.66 &  0.01 &  0.00 \\
      V(2)  &  0.66 & 0.08 &  0.00 &  0.67 &  0.05 &  0.00 &  0.66 &  0.01 &  0.00 \\
      V(3)  &  0.00 & 0.00 &  0.00 &  0.00 &  0.00 &  0.00 &  0.00 &  0.00 &  0.00 \\
      Sb(1) &  0.00 & 0.03 & -0.17 &  0.00 &  0.04 & -0.16 &  0.00 &  0.07 & -0.16 \\
      Sb(2) &  0.00 & 0.03 &  0.17 &  0.00 &  0.04 &  0.16 &  0.00 &  0.07 &  0.16 \\
      Sb(3) &  0.00 & 0.03 & -0.17 &  0.00 &  0.04 & -0.16 &  0.00 &  0.07 & -0.16 \\
      Sb(4) &  0.00 & 0.03 &  0.17 &  0.00 &  0.04 &  0.16 &  0.00 &  0.07 &  0.16 \\
      Sb(5) &  0.00 & 0.00 &  0.00 &  0.00 &  0.00 &  0.00 &  0.00 &  0.00 &  0.00   
    \end{tabular}
  \end{ruledtabular}
\end{table} 

The components of the displacement vector of the unstable phonon mode
at $L$ for the three compounds are given in Table~\ref{tab:disp-vec}.
As expected, the displacement vectors are qualitatively similar in all
three compounds.  I find that the nature of the displacement vector
does not change along the out-of-plane $M$--$L$ high-symmetry path,
except that the alkali ions are fixed at $M$ and for the presence of
complex phase factors for the wave vectors that do not lie at the
Brillouin zone corners. Within the $ab$ plane, the unstable mode
causes the V-V bonds orthogonal to the phonon propagation vector to
dimerize.  Additionally, the instability also causes out-of-plane
motion of Sb ions that do not lie in the kagome plane.



\begin{figure*}
  \includegraphics[width=\textwidth]{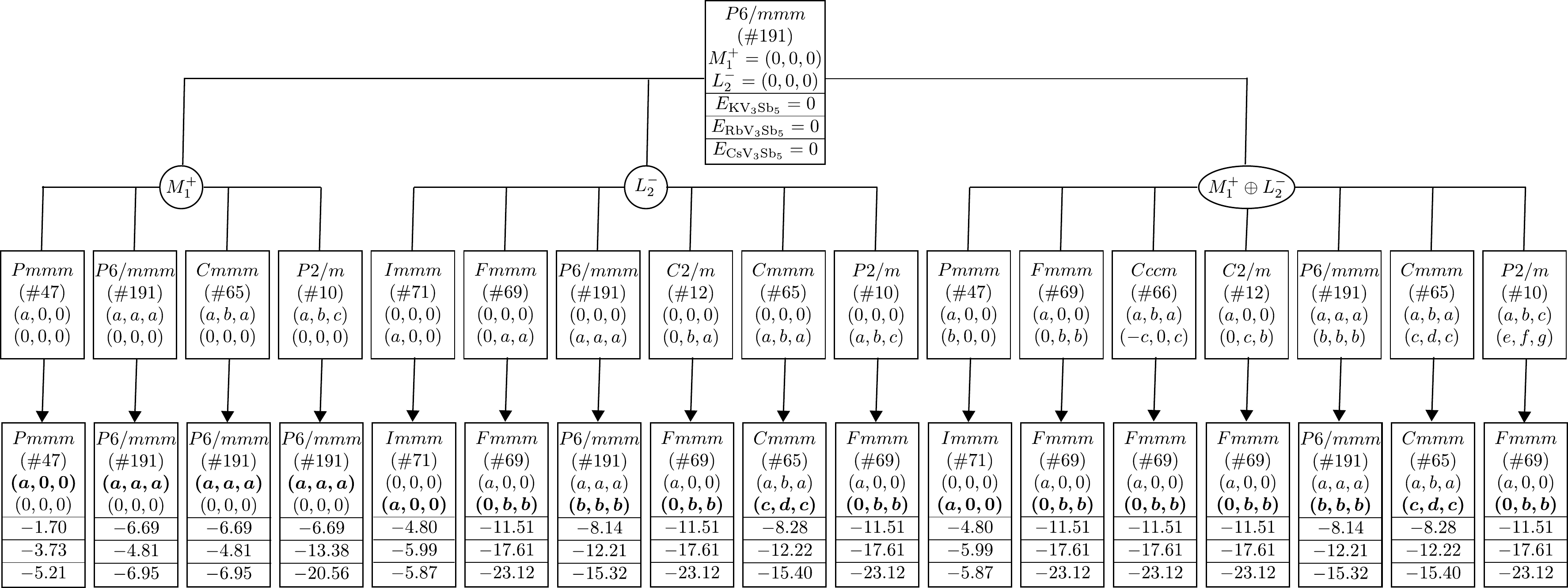}
  \caption{All possible isotropy subgroups and order parameters that
    can arise due to the $M^+_1$ and $L^-_2$ phonon instabilities of
    the $P6/mmm$ phase of the $A$V$_3$Sb$_5$ compounds. The 17
    possible structures enumerated in the boxes in the middle row were
    generated for all three compounds, and full structural relaxations
    were performed on them. The boxes in the bottom row shows the
    final structure after the relaxation.  Several low-symmetry
    initial structures relaxed to a higher-symmetry phase. For each
    phase, the space group symbol and number, order parameters of the
    $M^+_1$ and $L^-_2$ instabilities, and the total energy
    $E_{\textrm{$A$V$_3$Sb$_5$}}$ in meV per formula unit relative to
    the $P6/mmm$ phase are given in separate rows inside the boxes.
    The primary order parameter is given in bold face for the final
    structures. The structures with the same order parameter but
    different total energies are characterized by different atomic
    distances. The $Fmmm$ structure has the lowest energy in all three
    compounds.}
  \label{fig:isotr}
\end{figure*}

Although the unstable phonon branch in these materials is
non-degenerate at both $M$ and $L$, this instability can cause
distortion along all three V-V chains present in the kagome lattice
because the star of both $M$ $\left\{\left(0,\frac{1}{2},0\right),
\left(\frac{1}{2},0,0\right), \left(\frac{1}{2},\frac{1}{2},0\right)
\right\}$ and $L$ $\left\{\left(0,\frac{1}{2},\frac{1}{2}\right),
\left(\frac{1}{2},0,\frac{1}{2}\right),
\left(\frac{1}{2},\frac{1}{2},\frac{1}{2}\right) \right\}$ contain
three elements. As a consequence, the irreducible representations
(irreps) $M^+_1$ and $L^-_2$ of the unstable phonon modes at $M$ and
$L$, respectively, are three dimensional.  I used the {\sc isotropy}
software package to determine the distinct structural distortions that
are possible in the six-dimensional order parameter subspace defined
by the $M_1^+$ and $L_2^-$ instabilities.  I found seventeen
distortion vectors belonging to nine isotropy subgroups, and they are
shown in Fig.~\ref{fig:isotr}.  Using the calculated phonon
displacement vectors, I generated all seventeen possible distortions
corresponding to the isotropy subgroups on a $2\times2\times2$
supercell of the parent $P6/mmm$ phase and fully relaxed these
structures by minimizing both the atomic forces and lattice stresses.
 
In all three compounds, only distortions with the space groups $Immm$,
$Fmmm$, $P6/mmm$, $Cmmm$, and $Pmmm$ could be stabilized, and their
relative energies with respect to that of the undistorted phase are
shown in Fig.~\ref{fig:isotr}.  Interestingly, the relaxations
resulted in several $P6/mmm$ structures with different $c$-axis
periodicity and V-V distances that are characterized by dissimilar
total energies. The relative energetic rankings of the structures
belonging to different isotropy subgroups differ in the three
compounds.  However, the distorted structure with the space group
$Fmmm$ is lowest in all three compounds.  Symmetry-mode analysis using
the {\sc amplimodes} program shows that primary order parameter of
this phase is $(0,b,b)$ belonging to the the irrep $L^-_2$ and the
secondary order parameter is $(a,0,0)$ belonging to the irrep $M^+_1$,
which can also be surmised from the fact that the initial structure
with a $M_1^+$ $(a,0,0)$ distortion does not relax to the $Fmmm$
structure but the initial structure with a $L_2^-$ $(0,b,b)$
distortion develops additional $M_1^+$ $(a,0,0)$ distortion during
structural relaxation.  My calculations not only confirm Ratcliff
\textit{et al.}'s result that the gain in energy due to a combined
$M_1^+$ $(a,0,0)$ and $L_2^-$ $(0,b,b)$ $3Q$ distortion is larger than
due to either $M_1^+$ $(a,a,a)$ or $L_2^-$ $(b,b,b)$ $3Q$ distortion
\cite{ratc21}, they also show that the $Fmmm$ structure with the order
parameter $M_1^+$ $(a,0,0)$ $+$ $L_2^-$ $(0,b,b)$ has the lowest
energy among all possible structures due to the $M_1^+$ and $L_2^-$
instabilities.

The calculated structural parameters of the $Fmmm$ phase of the three
compounds are given in Tables \ref{tab:kfmmm}, \ref{tab:rfmmm}, and
\ref{tab:cfmmm} in the appendix.  The conventional
unit cell of this face-centered orthorhombic phase is related to that
of the parent hexagonal phase by the transformation matrix
 \[
    T = 
    \begin{pmatrix*}[r]
        2 & \phantom{-}2 & \phantom{-}0 \\
        2 & -2 & 0 \\
        0 & 0 & 2
    \end{pmatrix*}.
 \]
There are sixteen $A$V$_3$Sb$_5$ formula units per conventional unit
cell of the $Fmmm$ phase, involving repetitions of 2, 4 and 2 along
the $a$, $b$ and $c$ axes, respectively, with respect to the parent
phase.  The quadrupling along the $b$ axis also involves a shift of
the structural motif along the $a$ axis.  As a result, the structural
distortion cannot be described as a reconstruction of the
two-dimensional layers that preserve the angle between the lattice
vectors.

\begin{figure}
  \includegraphics[width=\columnwidth]{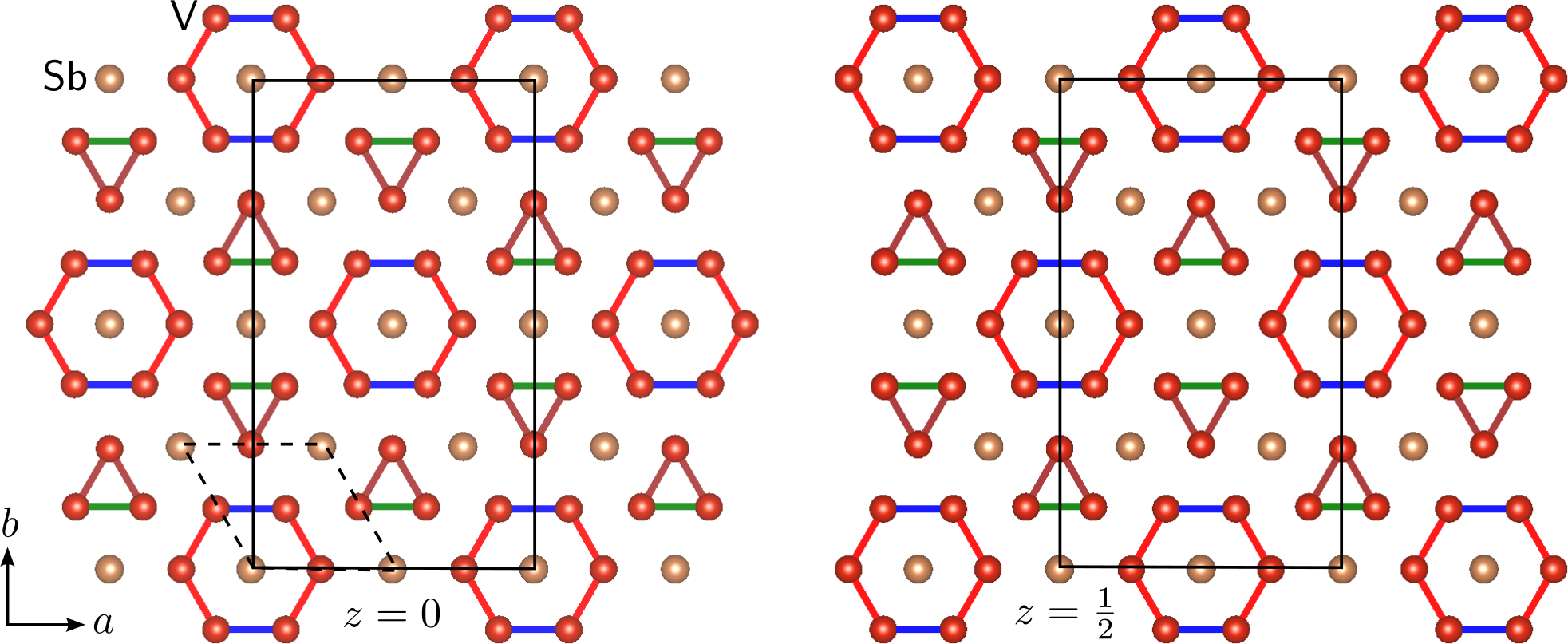}
  \caption{The $z = 0$ and $\frac{1}{2}$ kagome layers in the
    tri-hexagonal $Fmmm$ phase of $A$V$_3$Sb$_5$ compounds. The dashed
    and solid black lines enclose the planar unit cells of the $P6/mmm$ and
    $Fmmm$ phases, respectively. The V-V bonds of different lengths
    within the triangles and hexagons are depicted using different
    colors.  There are four different dimerized V-V bonds in the
    $Fmmm$ phase. }
  \label{fig:fmmmkag}
\end{figure}

The V$_3$Sb kagome plane in the $Fmmm$ phase is shown in
Fig.~\ref{fig:fmmmkag}.  
%
%
The colored solid lines in the figure indicate the V-V bonds that are
contracted relative to the parent phase and different colors denote
different bond lengths.  The distortions involve the formation of a
tri-hexagonal pattern (also called inverse Star of David) predicted
for the symmetric 3$Q$ condensation of the unstable mode at $M$ by Tan
\textit{et al.}\ \cite{tan21}, and the tri-hegaxonal pattern is
laterally shifted in the neighboring planes as predicted by Ratcliff
\textit{et al.}\ for the simultaneous $M_1^+$ $(a,0,0)$ and $L_2^-$
$(0,b,b)$ condensation of the unstable modes \cite{ratc21}.
Furthermore, I find that the both the V triangles and hexagons in the
kagome plane are made up of two inequivalent bond lengths.  One V triangle
alternates between the V hexagons along the $a$ axis, whereas two V
triangles alternate between the V hexagons along the $b$ axis of the
orthogonal structure.
Additionally, there are bucklings in the $A$ and Sb layers that lie
between the kagome sheets (not shown).

\begin{figure*}[t]
  \includegraphics[width=\textwidth]{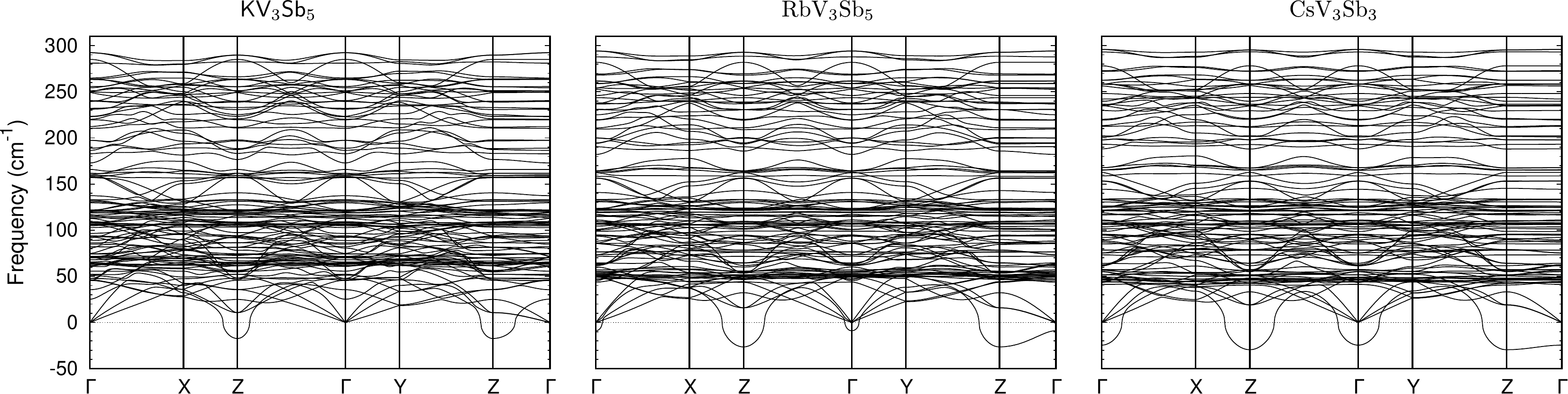}
  \caption{Calculated phonon dispersions of fully-relaxed
    KV$_3$Sb$_5$, RbV$_3$Sb$_5$, and CsV$_3$Sb$_5$ in the $Fmmm$ phase
    otained using the optB88-vdW functional show all three materials
    have an unstable branch with largest instability at $Z$.  The
    dispersions are plotted along the path $\Gamma$ $(0,0,0)
    \rightarrow$ $X$ $(1,0,0) \rightarrow$ $Z$ $(1,1,0) \rightarrow$
    $\Gamma$ $(0,0,0) \rightarrow$ $Y$ $(0,1,0) \rightarrow$ $Z$
    $(1,1,0) \rightarrow$ $\Gamma$ $(1,1,1)$.  The coordinates are
    given in terms of the conventional reciprocal lattice vectors.}
  \label{fig:phdfmmm}
\end{figure*}

\subsection{Is the $\bm{Fmmm}$ phase dynamically stable?}

Synchrotron x-ray diffraction data measured by Ortiz \textit{et
  al.}\ indicate a periodicity of four along the out-of-plane
direction in the low-temperature phase of CsV$_3$Sb$_5$
\cite{orti21b}. However, the $Fmmm$ structure discussed above only
exhibits a periodicity of two with respect to the $P6/mmm$ phase along
the $c$ axis.  I calculated the phonon dispersions of the three
compounds in the $Fmmm$ phase to investigate whether this phase shows
additional instability that leads to a $4c$ periodicity with respect
to the parent phase, and the results are shown in
Fig.~\ref{fig:phdfmmm}.  There is indeed an unstable branch in all
three compounds, and the largest instability occurs at the $Z$ point
$(0,0,1)$ with respect to the conventional lattice.  The instability
only occurs around $Z$ in KV$_3$Sb$_5$.  In RbV$_3$Sb$_5$ and
CsV$_3$Sb$_5$, this branch is unstable along the path
$\Gamma$-$Z$. The instability at $\Gamma$ is small in RbV$_3$Sb$_5$,
but it is almost as large as the instability at $Z$ in CsV$_3$Sb$_5$.
 
The $Z$ point of the $Fmmm$ structure is $A$ $(0,0,\frac{1}{2})$ in
terms of the reciprocal lattice of the parent $P6/mmm$
phase. Therefore, this instability does not quadruple the
periodicity along $c$.
 
The $Z$ point in the Brillouin zone of the $Fmmm$ structure has only
one element in its star, and the unstable mode is nondegenerate with
the irrep $Z_4^-$.  The only isotropy subgroup of this irrep is
$Cmcm$. So this instability causes a transition from a face-centered
orthorhombic to a base-centered orthorhombic structure. I used the
eigenvector of this unstable mode to generate 72-atom $Cmcm$
structures of all three compounds and performed full structural
relaxations.  The total energies of thus stabilized $Cmcm$ structures
of KV$_3$Sb$_5$, RbV$_3$Sb$_5$ and CsV$_3$Sb$_5$ are 0.3, 0.3 and 0.9
meV per formula unit lower than that of the respective $Fmmm$
structures.

\begin{figure}
  \includegraphics[width=\columnwidth]{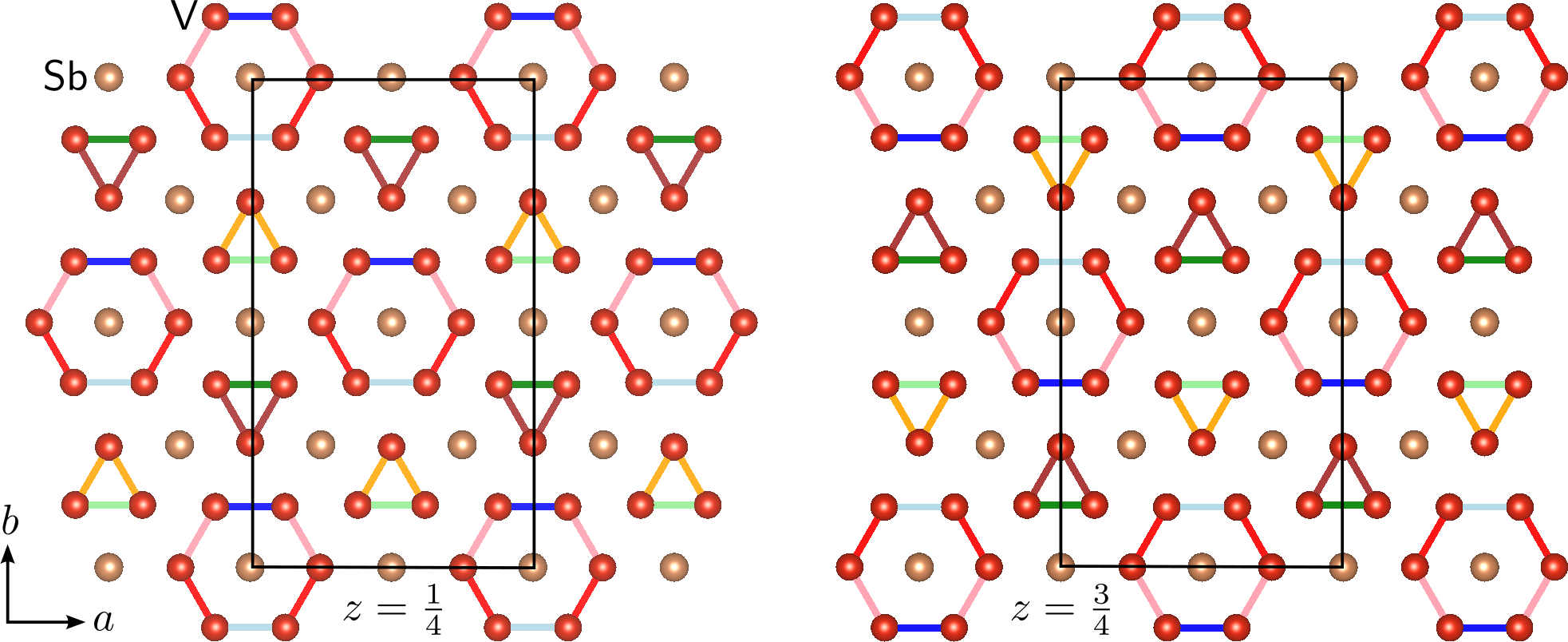}
  \caption{The kagome layers in the $Cmcm$ phase of $A$V$_3$Sb$_5$
    compounds. The V-V bonds of different lengths within the triangles
    and hexagons are depicted using different colors. There are eight
    different dimerized V-V bonds in the $Cmcm$ phase.  Individual
    kagome layers in this phase lack the mirror symmetry perpendicular
    to the $b$ axis, which
    is restored in the full lattice due to the presence of a glide
    plane.}
  \label{fig:cmcmkag}
\end{figure}

The calculated structural parameters of this phase for the three
compounds are given in Table \ref{tab:kcmcm}, \ref{tab:rcmcm}, and
\ref{tab:ccmcm} in the appendix.  Fig.~\ref{fig:cmcmkag} shows its kagome layer where
different V-V bond lengths of the triangles and hexagons are painted
with different colors.  Compared to the $Fmmm$ phase, the
nearest-neighbor V triangles in the $Cmmm$ phase become inequivalent
through alternate compression and expansion of their areas.  There is
additional differentiation of bonds in the V hexagons such that it is
now composed of four different bond lengths.  As a result, the
individual kagome layers lose the mirror symmetry that is
perpendicular to its $b$ axis, while the reciprocal lattice still
exhibits this mirror symmetry because of the presence of a glide plane
perpendicular to the $b$ axis in the $Cmmm$ phase.

The $Z_4^-$ instability of the $Fmmm$ phase corresponds to the $A_6^+$
irrep of the $P6/mmm$ phase.  Remarkably, there are no unstable modes
at $A$ $(0,0,\frac{1}{2})$ in the parent $P6/mmm$ phase. However, the
lowest-frequency mode at $A$ has the irrep $A_6^+$.  It has a small
frequency of 11 cm$^{-1}$ and exhibits a strong dispersion
characteristic of a soft mode.  Apparently, the transition to the
$Fmmm$ phase makes it unstable.  This mode is doubly degenerate and
structure-relationship analysis using the {\sc amplimodes} code shows
that the order parameter has the direction
$(\frac{1}{2}c,\frac{-\sqrt{3}}{2}c)$. Therefore, the $Cmcm$ phase can
be characterized as having the $4Q$ order parameter $M_1^+$ $(a,0,0)$
$+$ $L_2^-$ $(0,b,b)$ $+$ $A_6^+$
$(\frac{1}{2}c,\frac{-\sqrt{3}}{2}c)$.

\subsection{Are there any structures lower in energy than the $\bm{Cmcm}$ phase?}

The magnitudes of the imaginary frequencies of the unstable phonon
modes at $Z$ and $\Gamma$ in the $Fmmm$ phase of CsV$_3$Sb$_5$ are
comparable. So it is possible that there are other structures lower in
energy than the $Cmmm$ phase in this material.  The unstable mode at
$\Gamma$ has the irrep $\Gamma_4^-$.  It is nondegenerate, and its
only isotropy subroupg is $Fmm2$.  The isotropy subgroup due to
coupled $Z_4^-$ and $\Gamma_4^-$ instabilities is $Amm2$.  I generated
144-atom conventional unit cells of these structures for CsV$_3$Sb$_5$
and fully relaxed them.  The calculated total energies of the $Fmm2$
and $Amm2$ phases are 0.7 and 0.6 meV/f.u.\ higher than the $Cmcm$
phase, respectively.

Congnizant of Ortiz \textit{et al.}'s report of a $4c$ periodicity in
CsV$_3$Sb$_5$ \cite{orti21b}, I also studied the possibility that the
instability at $(0,0,\frac{1}{2})$, \textit{i.e.}\ the midpoint of the
$\Gamma$--$Z$ path of the $Fmmm$ phase, produces a lower energy
structure.  Since $(0,0,\frac{1}{2})$ and $(0,0,-\frac{1}{2})$ form a
star, there are three possible distortions due to this instability.
They correspond to a single-\textit{q},  double-\textit{q} with equal
magnitude, and  double-\textit{q} with unequal magnitude
condensations of the phonon instability. These structures have the
space group $Cmcm$ and exhibit $4c$ periodicity with respect to the
parent $P6/mmm$ phase.  The calculated total energies of these
structures are 0.5, 0.7, and 0.4 meV/f.u higher than that of the
$Cmcm$ phase due to the $Z$ instability, respectively.

In principle, the $Cmcm$ phase due to the $Z$ instability may also
exhibit phonon instabilities and might not be the lowest energy phase
of these materials.  The $Cmcm$ phase has 72 atoms per primitive unit
cell, and I did not have the computational resources to calculate its
full phonon dispersions even using the frozen phonon approach as it
requires at least a $2\times2\times2$ supercell.  However, frozen
phonon calculations on the smaller 144-atom $2\times1\times1$ and
$1\times1\times2$ supercells of CsV$_3$Sb$_5$ were feasible. These
yield accurate phonon frequencies at the wave vectors commensurate to
these supercells, and thus calculated phonon frequencies were all
real.  In particular, I did not find any instabilities that correspond
to a quadrupling of the hexagonal cell along the $c$ direction.
Therefore, the synchrotron x-ray diffraction experiment of Ortiz
\textit{et al.}\  showing a periodicity of $4c$ in CsV$_3$Sb$_5$ at low
temperatures remain a puzzle \cite{orti21b}.  One possibility that
could explain their observation is the presence of additional stacking
disorder due to the broken $m_y$ symmetry of the kagome planes in the
$Cmcm$ phase.  The occurrence of a substantial amount of 180$^\circ$
stacking faults can lead to peaks corresponding to a $4c$ periodicity
in the diffraction data.  It is worthwhile to note that such a
180$^\circ$ stacking fault is not possible for the $Fmmm$ structure.

\subsection{Electronic structure of the $\bm{P6/mmm}$, $\bm{Fmmm}$, and
  $\bm{Cmcm}$ phases of KV$_3$Sb$_5$}

\begin{figure}
  \includegraphics[width=\columnwidth]{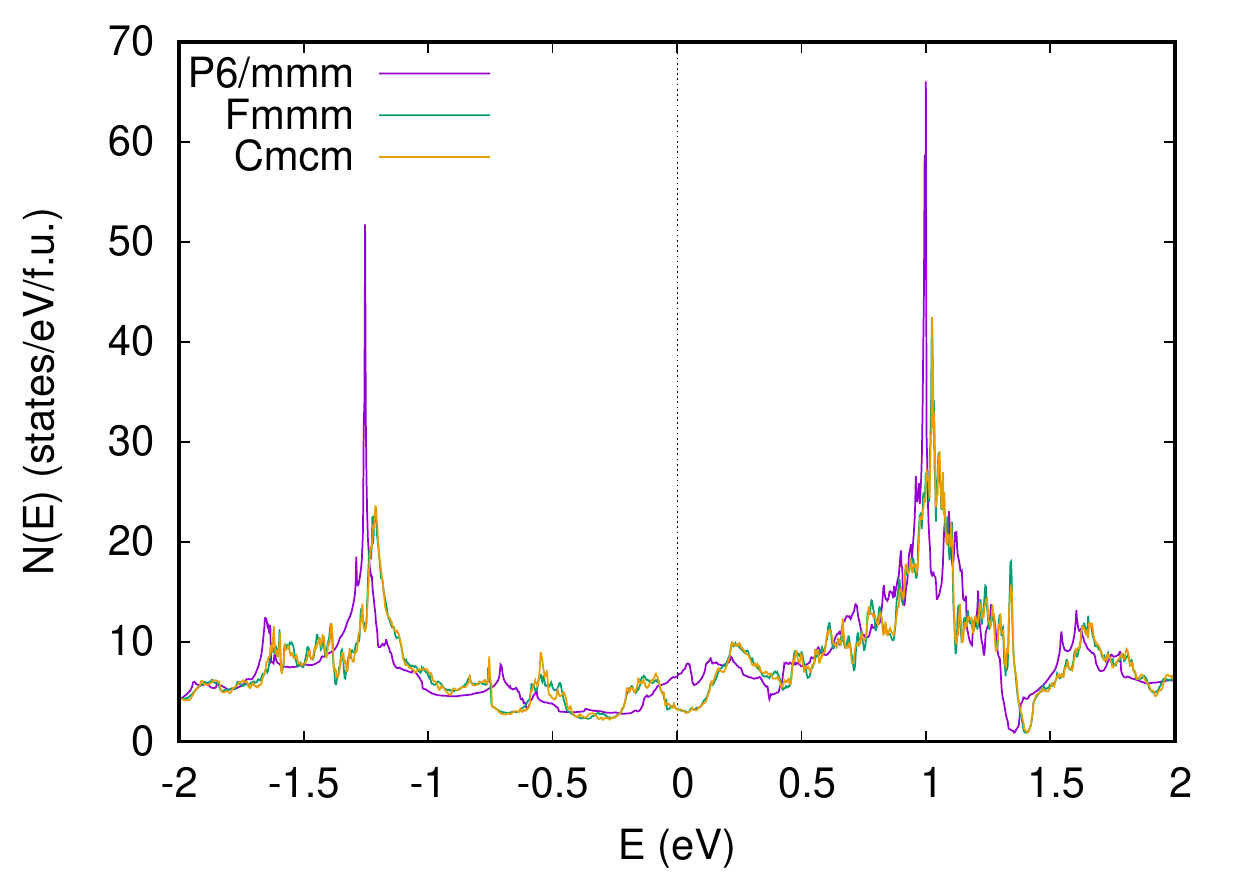}
  \caption{Calculated electronic density of states of KV$_3$Sb$_5$ in
    the $P6/mmm$, $Fmmm$, and $Cmcm$ phases.  }
  \label{fig:kvsdos}
\end{figure}

\begin{figure}[h]
  \includegraphics[width=\columnwidth]{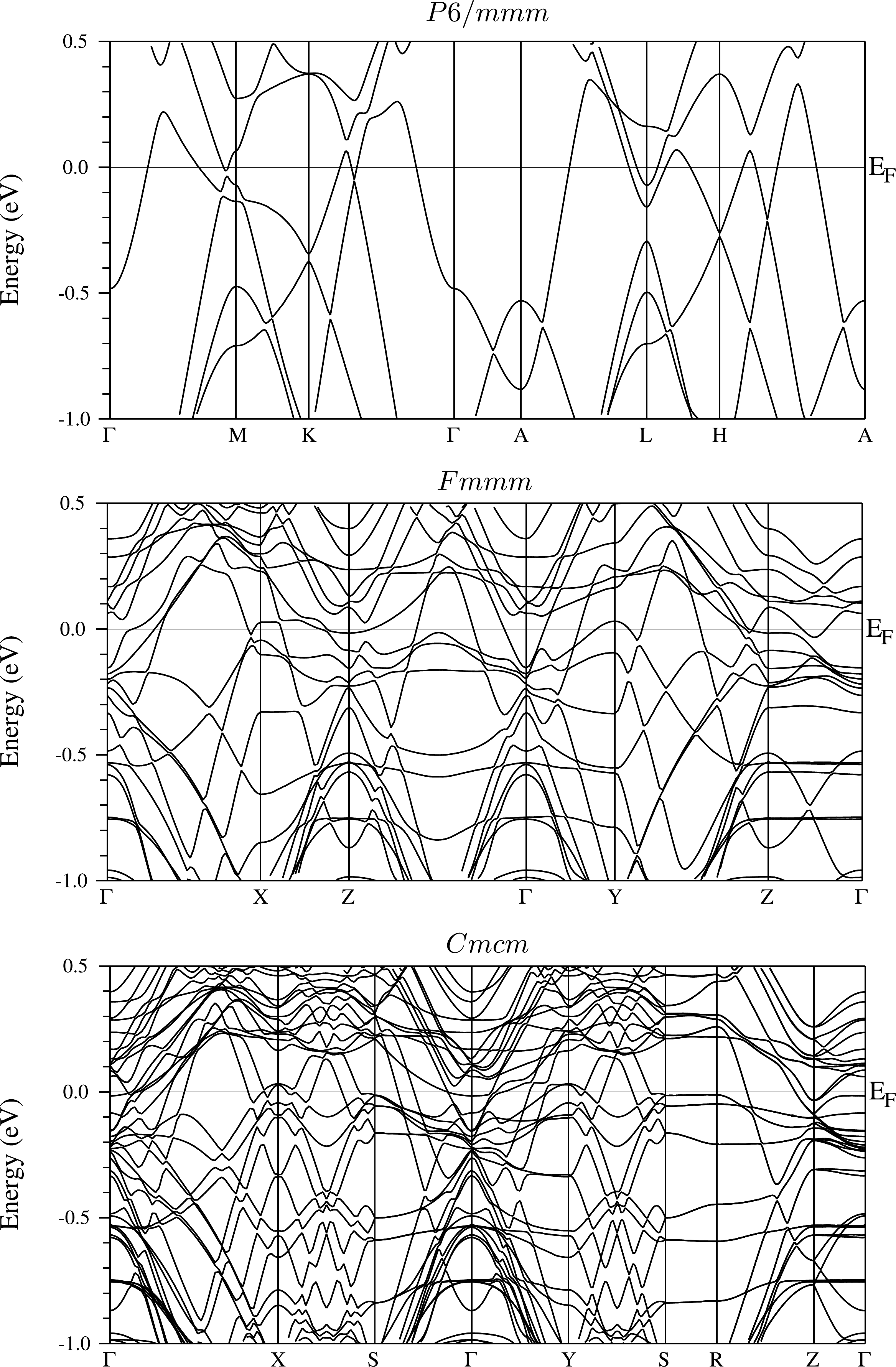}
  \caption{Calculated band structures of KV$_3$Sb$_5$ in the $P6/mmm$,
    $Fmmm$, and $Cmcm$ phases.  The high-symmetry points for the
    $Cmcm$ phase are $\Gamma$ $(0,0,0)$, $X$ $(1,0,0)$, $S$
    $(\frac{1}{2},\frac{1}{2},0)$, $Y$ $(0,1,0)$, $R$
    $(\frac{1}{2},\frac{1}{2},\frac{1}{2})$, and $Z$
    $(0,0,\frac{1}{2})$ in terms of the conventional reciprocal
    lattice vectors.}
  \label{fig:kvsbs}
\end{figure}

I now briefly discuss the electronic structure of KV$_3$Sb$_5$ in the
$P6/mmm$, $Fmmm$, and $Cmcm$ phases to illustrate how the structural
distortions modify the electronic properties.  Figs.~\ref{fig:kvsdos}
and \ref{fig:kvsbs} show the calculated electronic density of states
(DOS) and band structures of the three phases, respectively.  The band
structure of the parent $P6/mmm$ phase agrees with previous
calculations \cite{orti19,tan21,uyku21a,labo21,hu21b}, but the DOS
shows noticeable qualitative and quantitative differences.  In
particular, previous results show that the Fermi level lies near a
peak inside a valley. I instead find that the Fermi level lies on the
shoulder of a peak outside the valley. Regardless, the calculated DOS
at the Fermi level is relatively high with a value of 6.52
eV$^{-1}$ per formula unit both spin basis or 1.09 eV$^{-1}$/V per spin.

The structural distortions strongly modify the electronic structure
near the Fermi level.  The valley formed by two peaks near the Fermi
level widens and the Fermi level now lies near the bottom inside the
valley.  The DOS values are $\sim$3 eV$^{-1}$ per formula unit both
spin basis for the $Fmmm$ and $Cmcm$ phases.  The V-V bond distances
change by up to 5\% as the $P6/mmm$ phase distorts to the $Fmmm$
structure, and the almost 50\% concomitant decrease in the DOS shows
that lattice couples strongly to the electronic states near the Fermi
level. However, the electronic structure away from the Fermi level
also shows significant changes. For example, the pair of close-lying
peaks at $-0.71$ and $-0.56$ eV in the $P6/mmm$ phase get
reconstructed to a broader valley.  Therefore, the structural
transition from the $P6/mmm$ to $Fmmm$ phase cannot merely be ascribed
to a Peierls instability and also has characteristics of a V-V bonding
instability. On the other hand, the V-V distances change by at most
$0.80\%$ in the course of the $Fmmm$-to-$Cmcm$ structural transition,
and these structures have the same DOS value within the errors due to
the coarseness of the $k$-point grid used in the calculations.  In
fact, the changes in the electronic structure away from the Fermi
level are more noticeable even if they are relatively small.  So this
transition is mainly due to V-V bonding instability.

The band structure of the parent $P6/mmm$ phase has three saddle
points near the Fermi level at the $M$ point.  There are additional
saddle points near $-0.7$ eV at $M$, $-0.5$ eV at $\Gamma$, and $0.2$
and $-0.7$ eV at $L$.  If one examines the folded bands at $\Gamma$ in
the $Fmmm$ phase, one can see that the van Hove singularities just
below the Fermi level at $M$ in the $P6/mmm$ phase get strongly
reconstructed.  However, the van Hove singularities just above the
Fermi level and at $-0.7$ eV persist at shifted positions.  This
suggest that CDW formation involves strong scattering mainly among the
two saddle points just below the Fermi level.  In the $P6/mmm$ phase
there are two electron band near the Fermi level around $L$, and only
the one that forms smaller Fermi sheet gets completely reconstructed
$\Gamma$ in the $Fmmm$ phase.  This shows scattering within van Hove
singularities dos not fully capture the mechanism of CDW formation.
The other electron band at $L$ survives when it is folded to $\Gamma$,
and such an electron band has been observed in ARPES experiments
\cite{luo21a}.  Interestingly, both the $Fmmm$ and $Cmcm$ phases
exhibit saddle points close to the Fermi level, which implies that
there are van Hove singularities near the Fermi level even in low
symmetry phases.  This is consistent with ARPES experiments on the
low-temperature phase that observe emergence of new van Hove
singularities \cite{cho21}.

\section{Summary and Conclusions}

In summary, I have explored the ground state structures of
KV$_3$Sb$_5$, RbV$_3$Sb$_5$, and CsV$_3$Sb$_5$ guided by the presence
of phonon instabilities present in these materials. I used
group-theoretical analysis to identify the seventeen different
structural distortions that are possible due to the $M_1^+$ and
$L_2^-$ phonon instabilities in the parent $P6/mmm$ phase.  I
generated these structures for the three compounds and studied their
energetics by performing full structural relaxations.  The $Fmmm$
phase with the order parameter $M_1^+$ $(a,0,0)$ $+$ $L_2^-$ $(0,b,b)$
has the lowest energy in all three compounds among these seventeen
possibilities. However, the three compounds in this phase exhibits a
dynamical instability at the $Z$ $(0,0,1)$ point, which corresponds to
the $A$ $(0,0,\frac{1}{2})$ point of the parent $P6/mmm$ phase.
Condensation of this instability leads to a base-centered orthorhombic
structure with the space group $Cmcm$ and $4Q$ order parameter
$M_1^+$ $(a,0,0)$ $+$ $L_2^-$ $(0,b,b)$ $+$ $A_6^+$
$(\frac{1}{2}c,\frac{-\sqrt{3}}{2}c)$.

\section{acknowledgement}
This work was supported by 
by GENCI-CINES (grant A0090911099) and the Swiss National
Supercomputing Center (grant s820).

\section{appendix}

Calculated lattice parameters and atomic positions of KV$_3$Sb$_5$,
RbV$_3$Sb$_5$, and CsV$_3$Sb$_5$ in the $Fmmm$ phase are given in
Tables~\ref{tab:kfmmm}, \ref{tab:rfmmm}, and \ref{tab:cfmmm},
respectively.  Those for the $Cmcm$ phase are give in
Tables~\ref{tab:kfmmm}, \ref{tab:rfmmm}, and \ref{tab:cfmmm},
respectively.  These were obtained using the optB88-vdw
exchange-correlation functional.

\begin{table}[h]
  \caption{\label{tab:kfmmm} Calculated atomic coordinates of
    KV$_3$Sb$_5$ in the $Fmmm$ phase. Calculated lattice
    parameters are $a = 10.96216, b = 18.99453$, and $c = 17.88179$ \AA.}
  \begin{ruledtabular}
    \begin{tabular}{lcccc}
      atom & Wyckoff pos. & $x$ & $y$ & $z$ \\
      \hline
      K    & $8i$  & 0        & 0        & 0.24866 \\
      K    & $8f$  & 1/4      & 1/4      & 1/4 \\
      V1   & $16o$ & 0.38111  & 0.12712  & 0 \\ 
      V2   & $16o$ & 0.62362  & 0.37688  & 0 \\
      V3   & $8g$  & 0.24645  & 0        & 0 \\
      V4   & $8h$  & 0        & 0.74560  & 0 \\
      Sb1  & $16m$ & 0        & 0.16668  & 0.63035 \\ 
      Sb2  & $32p$ & 0.75060  & 0.41652  & 0.12735 \\
      Sb3  & $16m$ & 0        & 0.66700  & 0.62709 \\ 
      Sb4  & $4b$  & 0        & 0        & 1/2 \\
      Sb5  & $8e$  & 1/4      & 1/4      & 0 \\
      Sb6  & $4a$  & 0        & 0        & 0
    \end{tabular}
  \end{ruledtabular}
\end{table}

\begin{table}
  \caption{\label{tab:rfmmm} Calculated atomic coordinates of
    RbV$_3$Sb$_5$ in the $Fmmm$ phase. Calculated lattice
    parameters are $a = 10.98970, b = 19.03729$, and $c = 18.19619$ \AA.}
  \begin{ruledtabular}
    \begin{tabular}{lcccc}
      atom & Wyckoff pos. & $x$ & $y$ & $z$ \\
      \hline
      Rb   & $8i$  & 0        & 0        & 0.25132 \\
      Rb   & $8f$  & 1/4      & 1/4      & 1/4 \\
      V1   & $16o$ & 0.12305  & 0.37708  & 0 \\ 
      V2   & $16o$ & 0.88188  & 0.12732  & 0 \\
      V3   & $8g$  & 0.74589  & 0        & 0 \\
      V4   & $8h$  & 0        & 0.24532  & 0 \\
      Sb1  & $16m$ & 0        & 0.83286  & 0.62440 \\ 
      Sb2  & $32p$ & 0.75080  & 0.08356  & 0.12452 \\
      Sb3  & $16m$ & 0        & 0.33334  & 0.62788 \\ 
      Sb4  & $4b$  & 0        & 0        & 1/2 \\
      Sb5  & $8e$  & 1/4      & 1/4      & 0 \\
      Sb6  & $4a$  & 0        & 0        & 0
    \end{tabular}
  \end{ruledtabular}
\end{table}

\begin{table}
  \caption{\label{tab:cfmmm} Calculated atomic coordinates of
    CsV$_3$Sb$_5$ in the $Fmmm$ phase. Calculated lattice
    parameters are $a = 11.01874, b = 19.08467$, and $c = 18.69718$ \AA.}
  \begin{ruledtabular}
    \begin{tabular}{lcccc}
      atom & Wyckoff pos. & $x$ & $y$ & $z$ \\
      \hline
      Cs   & $8i$  & 0        & 0        & 0.75120 \\
      Cs   & $8f$  & 1/4      & 1/4      & 1/4 \\
      V1   & $16o$ & 0.38247  & 0.12750  & 0 \\ 
      V2   & $16o$ & 0.62265  & 0.37734  & 0 \\
      V3   & $8g$  & 0.24530  & 0        & 0 \\
      V4   & $8h$  & 0        & 0.74501  & 0 \\
      Sb1  & $16m$ & 0        & 0.33270  & 0.62051 \\ 
      Sb2  & $32p$ & 0.25102  & 0.08364  & 0.12056 \\
      Sb3  & $16m$ & 0        & 0.83336  & 0.62418 \\ 
      Sb4  & $4b$  & 0        & 0        & 1/2 \\
      Sb5  & $8e$  & 1/4      & 1/4      & 0 \\
      Sb6  & $4a$  & 0        & 0        & 0
    \end{tabular}
  \end{ruledtabular}
\end{table}

\begin{table}
  \caption{\label{tab:kcmcm} Calculated atomic coordinates of
    KV$_3$Sb$_5$ in the $Cmcm$ phase. Calculated lattice parameters
    are $a = 10.96187, b = 18.99383$, and $c = 17.88307$ \AA.}
  \begin{ruledtabular}
    \begin{tabular}{lcccc}
      atom & Wyckoff pos. & $x$ & $y$ & $z$ \\
      \hline
      K    & $8e$  & 0.25016    & 0         & 0 \\ 
      K    & $8f$  & 0          & 0.25006   & 0.50138 \\
      V1   & $8g$  & 0.11920    & 0.12292   & 1/4 \\
      V2   & $8g$  & 0.11864    & 0.37716   & 1/4 \\
      V3   & $8g$  & 0.37717    & 0.37333   & 1/4 \\
      V4   & $8g$  & 0.37552    & 0.12708   & 1/4 \\
      V5   & $4c$  & 0          & 0.00421   & 1/4 \\
      V6   & $4c$  & 0          & 0.49543   & 1/4 \\
      V7   & $8g$  & 0.25351    & 0.24950   & 1/4 \\
      Sb1  & $16h$ & 0.24928    & 0.16648   & 0.37757 \\
      Sb2  & $16h$ & 0.74951    & 0.16656   & 0.87713 \\
      Sb3  & $8f$  & 0          & -0.08294  & 0.37732 \\
      Sb4  & $8f$  & 0          & 0.41694   & 0.87684 \\
      Sb5  & $8f$  & 0          & 0.41667   & 0.38043 \\
      Sb6  & $8f$  & 0          & -0.08332  & 0.88024 \\
      Sb7  & $8g$  & 0.24992    & -0.00003  & 1/4 \\
      Sb8  & $4c$  & 0          & 0.74994   & 1/4 \\
      Sb9  & $4c$  & 0          & 0.24991   & 1/4
    \end{tabular}
  \end{ruledtabular}
\end{table}

\begin{table}
  \caption{\label{tab:rcmcm} Calculated atomic coordinates of
    RbV$_3$Sb$_5$ in the $Cmcm$ phase. Calculated lattice parameters
    are $a = 10.98897, b = 19.03705$, and $c = 18.19831$ \AA.}
  \begin{ruledtabular}
    \begin{tabular}{lcccc}
      atom & Wyckoff pos. & $x$ & $y$ & $z$ \\
      \hline
      Rb   & $8e$  & 0.25015    & 0         & 0 \\ 
      Rb   & $8f$  & 0          & 0.25006   & 0.50134 \\
      V1   & $8g$  & 0.11848    & 0.12275   & 1/4 \\
      V2   & $8g$  & 0.11781    & 0.37737   & 1/4 \\
      V3   & $8g$  & 0.37800    & 0.37323   & 1/4 \\
      V4   & $8g$  & 0.37585    & 0.12736   & 1/4 \\
      V5   & $4c$  & 0          & 0.00445   & 1/4 \\
      V6   & $4c$  & 0          & 0.49513   & 1/4 \\
      V7   & $8g$  & 0.25405    & 0.24932   & 1/4 \\
      Sb1  & $16h$ & 0.24904    & 0.16639   & 0.37479 \\
      Sb2  & $16h$ & 0.74935    & 0.16650   & 0.87422 \\
      Sb3  & $8f$  & 0          &-0.08278   & 0.37468 \\
      Sb4  & $8f$  & 0          & 0.41705   & 0.87409 \\
      Sb5  & $8f$  & 0          & 0.41665   & 0.37797 \\
      Sb6  & $8f$  & 0          &-0.08333   & 0.87775 \\
      Sb7  & $8g$  & 0.24989    & 0.00000   & 1/4 \\
      Sb8  & $4c$  & 0          & 0.74994   & 1/4 \\
      Sb9  & $4c$  & 0          & 0.24989   & 1/4 \\
    \end{tabular}
  \end{ruledtabular}
\end{table}

\begin{table}
  \caption{\label{tab:ccmcm} Calculated atomic coordinates of
    RbV$_3$Sb$_5$ in the $Cmcm$ phase. Calculated lattice parameters
    are $a = 11.01910, b = 19.08816$, and $c = 18.69497$ \AA.}
  \begin{ruledtabular}
    \begin{tabular}{lcccc}
      atom & Wyckoff pos. & $x$ & $y$ & $z$ \\
      \hline
      Cs1  & $8e$  & 0.25043   &   0  & 0 \\
      Cs2  & $8f$  & 0         &   0.25015  & 0.50116 \\
      V1   & $8g$  & 0.11862   &   0.12276  & 1/4 \\
      V2   & $8g$  & 0.11650   &   0.37773  & 1/4 \\
      V3   & $8g$  & 0.38079   &   0.37378  & 1/4 \\
      V4   & $8g$  & 0.37371   &   0.12831  & 1/4 \\
      V5   & $4c$  & 0         &   0.00431  & 1/4 \\
      V6   & $4c$  & 0         &   0.49437  & 1/4 \\
      V7   & $8g$  & 0.25450   &   0.24772  & 1/4 \\
      Sb1  & $16h$ & 0.24849   &   0.16621  & 0.37153 \\
      Sb2  & $16h$ & 0.74953   &   0.16655  & 0.86964 \\
      Sb3  & $8f$  & 0         &  -0.08241  & 0.37150 \\
      Sb4  & $8f$  & 0         &   0.41695  & 0.86956 \\
      Sb5  & $8f$  & 0         &   0.41662  & 0.37448 \\
      Sb6  & $8f$  & 0         &  -0.08335  & 0.87384 \\
      Sb7  & $8g$  & 0.24962   &   0.00003  & 1/4 \\
      Sb8  & $4c$  & 0         &   0.74989  & 1/4 \\
      Sb9  & $4c$  & 0         &   0.24967  & 1/4 \\
    \end{tabular}
  \end{ruledtabular}
\end{table}

\end{document}